\documentclass{elsart}

\usepackage{natbib}

\usepackage{graphicx}

\usepackage{amssymb}

\begin{document}

\begin{frontmatter}


\title{Goodness-of-fit tests of Gaussianity: constraints on the cumulants of the MAXIMA data}
\author[ifca,fism]{A.M. Aliaga},
\author[ifca]{E. Mart\'{\i}nez-Gonz\'alez},
\author[ifca,purd]{L. Cay\'on},
\author[ovie]{F. Arg\"ueso},
\author[ifca]{J.L. Sanz},
\author[ifca]{R.B. Barreiro}

\address[ifca]{Instituto de F\'{\i}sica de Cantabria, CSIC--Univ. de Cantabria, Avda. los Castros s/n, 39005 Santander, Spain}
\address[fism]{Dpto. de F\'{\i}sica Moderna. Univ. de Cantabria, Avda. Los Castros s/n, 39005 Santander, Spain}
\address[purd]{Physics Department, Purdue University, 525 Northwestern Avenue, West Lafayette, IN 47907-2036, USA}
\address[ovie]{Dpto. de Matem\'aticas, Univ. de Oviedo, C/ Calvo Sotelo s/n, 33007 Oviedo, Spain}


\begin{abstract}

In this work, goodness-of-fit tests are adapted and applied to CMB maps to detect possible non-Gaussianity. We use Shapiro-Francia test and two Smooth goodness-of-fit tests: one developed by Rayner and Best and another one developed by Thomas and Pierce. The Smooth tests test small and smooth deviations of a prefixed probability function (in our case this is the univariate Gaussian). Also, the Rayner and Best test informs us of the kind of non-Gaussianity we have: excess of skewness, of kurtosis, and so on. These tests are optimal when the data are independent. We simulate and analyse non-Gaussian signals in order to study the power of these tests. These non-Gaussian simulations are constructed using the Edgeworth expansion, and assuming pixel-to-pixel independence. As an application, we test the Gaussianity of the MAXIMA data. Results indicate that the MAXIMA data are compatible with Gaussianity. Finally, the values of the skewness and kurtosis of MAXIMA data are constrained by $|S| \le 0.035$  and $|K| \le 0.036$ at the 99$\%$ confidence level.

\end{abstract}


\end{frontmatter}

\section{Introduction}

Standard inflationary theories establish that primordial density fluctuations in the Universe had a Gaussian distribution. These fluctuations grew because of the gravitational force and created the structures we see today in the Universe. These fluctuations also left their imprint in the cosmic microwave background (CMB) as primordial anisotropies, which should also follow a Gaussian distribution. Therefore, if we test CMB Gaussianity, we are testing Gaussianity of primordial fluctuations and the validity of Standard Inflation. We can do the study of Gaussianity in several spaces: real space, Fourier space and wavelet space. In this paper we work in real space and apply three goodness-of-fit tests. We study their power to distinguish between Gaussian and non-Gaussian maps. As an application, we test the Gaussianity of the MAXIMA data. \citet{cayon} find these data compatible with Gaussianity. We will refer to that paper for most of the goodness-of-fit application to the MAXIMA data. In the present paper we give constraints on the skewness and kurtosis of these data. 

The organization of the paper is as follows. Goodness-of-fit tests are presented and tested in Section 2. Section 3 is dedicated to analyse the MAXIMA data and to constrain their skewness and kurtosis values. Finally, Section 4 is dedicated to discussion and conclusions. 

\section{Goodnes-of-fit statistics}

Given a sample of uncorrelated and normalized CMB data, we want to answer the question: ``how well the data agree with the population of a Gaussian distribution $N(0,1)$?''. 

\subsection{Shapiro-Francia test}

There are many goodness-of-fit methods to test Gaussianity (for a review see D'Agostino and Stephens, 1986). The Shapiro and Francia test is one of these methods (Shapiro and Francia, 1972). The statistic associated to this test study the correlation between a Gaussian distribution and our experimental data. We estimate a one-dimensional array $\vec c$ corresponding to the expected sorted values obtained from independent Gaussian simulations $N(0,1)$. Then we define $\vec b = \vec c / || \vec c ||$, where $|| \vec c ||^2 = \sum_i c_i^2$. Given our sorted data $\vec x$, the Shapiro-Francia statistic $SF$ is defined as follows:

\begin{displaymath}
SF=\frac{1}{n \sigma^2}\bigg(\sum_{i=1}^n b_i x_i \bigg)^2,
\end{displaymath}

where $n$ is the number of data and $\sigma$ is its dispersion. Note that, if $\vec x = \vec c \cdot \sigma $, as expected in the Gaussian case, then $SF \approx 1$. Thus, deviations from Gaussianity will result in values smaller than one, because, in this case, the correlation between $\vec x$ and $\vec c$ is smaller than for the Gaussian case.  

\subsection{Smooth tests}

Smooth tests are constructed to discriminate between a predetermined function $f(x)$ and a second one that deviates smoothly from the former. Given a statistical variable $x$ and $n$ independent realizations $(x_1, \ldots, x_n) \equiv \vec x$, we want to test if the probability function of $x$ is equal to $f(x)$ (in our case $N(0,1)$). We consider an alternative probability density function $f(x,\theta)$ (where $\theta$ is a parameter vector) that deviates smoothly from $f(x)$ and $f(x,\theta_0) = f(x)$ (we consider $\theta_0 =0$). In other words, we want to test the \emph{null hypothesis}: $\theta = \theta_0$.

The Smooth tests we are going to consider are based on the so-called \emph{score statistic} (these tests are widely explained in Cox and Hinkley, 1974). One defines the natural logarithm of the likelihood as $\ell (\vec x, \theta) \equiv \sum_{i=1}^n \log f(x_i, \theta)$, the vector $U$ of components  $U_i(\theta) \equiv \partial \ell(\vec x, \theta) / \partial \theta_i$ and the matrix $I$ of components $I_{ij}(\theta)\equiv \langle U_i(\theta)U_j(\theta) \rangle = - \langle \partial^2 \ell(\vec x, \theta) / \partial \theta_i \partial \theta_j \rangle$. The score statistic is closely related to the natural logarithm of the likelihood ratio and is given by $S = U^T(\theta_0) I^{-1}(\theta_0) U(\theta_0)$, where $U^T$ is the transpose vector of $U$. The null hypothesis is rejected for large values of $S$.

\subsubsection{Rayner-Best test}

These authors define an alternative probability density function of orden $k$ (\citet{rayner}) given by

\begin{displaymath}
g_k(x)= C(\theta_1,\ldots,\theta_k) \exp \bigg\{ \sum_{i=1}^k \theta_i h_i(x) \bigg\} f(x) .
\end{displaymath}

The $h_i$ functions are orthonormal on $f$ and $h_0(x)=1$. $C$ is a normalization constant. Then, the score statistic associated to the $k$ alternative is given by

\begin{displaymath}
S_k = \sum_{i=1}^k U_i^2 \qquad \textrm{with} \qquad U_i=\frac{1}{\sqrt n} \sum_{j=1}^n h_i(x_j).
\end{displaymath}

When Gaussianity is tested, then $h_n(x)=P_n(x)/s_n$, with $s_n=\sqrt{n!}$ and $P_0(x)=1$, $P_1(x)=x$ and for $n \geqslant 1$: $P_{n+1}(x)=xP_n(x)-nP_{n-1}(x)$ (Hermite-Chebishev polynomials). The statistic $S_k$ is related to cumulants of order $\leqslant k$: $ S_1  = n (\hat{\mu}_1)^2$, $S_2  = S_1 + n (\hat{\mu}_2 - 1 )^2 /2 $, $S_3  = S_2 + n (\hat{\mu}_3 - 3 \hat{\mu}_1 )^2 /6$, $S_4  =  S_3 + n ( [\hat{\mu}_4 - 3 ] - 6 [\hat{\mu}_2 -1] )^2  /24 $  and $S_5  =  S_4 + n ( \hat{\mu}_5 - 10 \hat{\mu}_3 + 15 \hat{\mu}_1)^2 /120 $, where $\hat{\mu}_{\alpha}=(\sum_{j=1}^n x_j^{\alpha})/n$. 

This test is directional, that is, it indicates how the actual distribution deviates from Gaussianity. For example, if $S_1$ and $S_2$ are small and $S_3$ is large, then the data have a large $\hat{\mu}_3$ value and also have a large skewness value.

When $n \rightarrow \infty$, the $S_k$ statistic is distributed as a $\chi_k^2$. This holds because if $n \rightarrow \infty$ then $U_i$ is Gaussian distributed (sum of a large number of independent variables).

\subsubsection{Thomas-Pierce test}

Thomas and Pierce (1979) take the cumulative distribution function of a $N(0,1)$ variable $x$: $y(x)= \textrm{erf}(x)$, where erf denotes the error function. If $x$ is Gaussian, then $y$ must be uniform distributed on the interval $[0,1]$, and the alternative probability function of $y$ is given by

\begin{displaymath}
g_k(x)=\exp \bigg\{ \sum_{i=1}^k \theta_i y^i - K(\theta)\bigg\},
\end{displaymath} 

where $K(\theta)$ is a normalization constant. The statistic score is then (with notation of Thomas and Pierce, 1979) ($k=1,2,3, \ldots$):

\begin{displaymath}
W_k=\sum_{i=1}^k \bigg( \sum_{j=1}^i a_{ij} u_j \bigg)^2 \quad, \quad 
u_j= \frac{1}{\sqrt{n}} \sum_{r=1}^n \bigg( y^j(x_r)- \frac{1}{1+j} \bigg).
\end{displaymath}

The coeficients $a_{ij}$ are given in Table 3 of \citet{thomas} (e. g. $a_{11}= 16.3172$, $a_{21}=-a_{22}=-27.3809$).

As it happens in the Rayner and Best test, when $n \rightarrow \infty$, the $W_k$ statistic is distributed as a $\chi_k^2$. Note that $\langle y^j \rangle = (1+j)^{-1}$, so this method is directional if we work with $y$ variable. 
 
\subsection{Gaussian simulations}

In \citet{cayon} the distributions of the previous statistics are calculated. These distributions are obtained for 50000 independent Gaussian simulations of maps with 2164 pixels (this number fits the number of central pixels in the MAXIMA map, later selected for analysis). The pixels are independent pixel-to-pixel. The data are normalized to zero mean and unit variance before tests are applied. Plots of these distributions can be found in \citet{cayon}.

\subsection{Non-Gaussian simulations. Edgeworth expansion.}

As an example of how well these methods work on discriminating between Gaussian and non-Gaussian data, we analyse simulated non-Gaussian maps obtained through the Edgeworth expansion (Mart\'{\i}nez-Gonz\'alez, 2002). The Edgeworth expansion allows to construct a distribution which has small deviations from Gaussianity, with desired values of skewness and kurtosis or any other cumulants. Given the Gaussian distribution $G(x)$, one denotes the cumulant of order $n$ by $k_n$, then, for small values of these cumulants, one can construct the density function $f(x)$:

\begin{displaymath}
f(x)=G(x) \bigg\{ 1+ \sum_{n=3}^\infty \frac{k_n}{n! 2^{n/2}} H_n(x/\sqrt{2}) + O(k_n k_n') \bigg\},
\end{displaymath} 

where $H_n$ is the Hermite polynomial of degree $n$, and $k_n$ is the cumulant of degree $n$. In particular the skewness is $k_3$ and the kurtosis is $k_4$. If we set all cumulants to zero except one, $f$ may not be positive definite and be not normalized. However, for small deviations (small values of the cumulants), one can set to zero the negative values of $f$ and then renormalize it, without disturbing the non-zero cumulants appreciably.  

We can use skewness ($S$) and kurtosis ($K$) as statistics. Given injected values of $S$ and $K$ (input values) we construct simulations and their $S$ and $K$ distributions. We compare these distributions with distributions of $S$ and $K$ for Gaussian simulations. The power of these statistics is given in Table \ref{tab0}. This table also show that, as mentioned before, the cumulants do not change significantly after setting to zero the negative values of the probability and its renormalization.

\begin{table}
  \begin{center}
  \caption{Average and dispersion for the skewness and 
kurtosis values obtained from 10000 simulations drawn from 
Edgeworth expansions assuming skewness and kurtosis
values denoted by S$\&$k(in). The power P of these two statistics is also
given in columns 4 and 5.}
  \label{tab0}
  \begin{tabular}{ccccc}\hline
  S$\&$K(in)& Mean/Disp (S)& Mean/Disp (K) & P(95/99$\%$) (S)& P(95/99$\%$) (K)\cr
  \hline

	0.0$\&$0.4&5.95e-4/0.0607&0.3170/0.1205&7.86/2.13&87.85/65.57\cr
	0.1$\&$0.0&0.0965/0.0503&-0.0342/0.0938&57.51/28.36&1.65/0.19\cr
	0.1$\&$0.9&0.1030/0.0780&0.7634/0.1524&57.49/32.25&67.26/37.05\cr
	0.3$\&$0.5&0.2949/0.0628&0.4179/0.1490&56.22/31.6&87.72/65.19\cr

  \hline
  \end{tabular}
  \end{center}
\end{table}


Over the same number of simulations with prefixed values of skewness and kurtosis, we calculate the power of the tests presented in this paper. The power for the $S_k$ statistic is shown in the Table \ref{tab1}. The power of the $W_k$ and $SF$ statistics is shown, respectively, in the Tables \ref{tab2} and \ref{tab3}. We can see that most of the presented tests have more power than the directly calculated skewness and kurtosis. The $W_2$ statistic seems to be the best discriminator in most of the cases.


\begin{table}
  \begin{center}
  \caption{Power at $95\%$ and $99\%$ confidence level for the $S_k$
statistics (notation used for the table $95\%/99\%$). Results based on 10000
 simulations Gaussian and non Gaussian simulations.
The non Gaussian ones were obtained from the Edgeworth expansion 
for different values of skewness and kurtosis S$\&$K(input).}
  \label{tab1}
  \begin{tabular}{ccccc}\hline
  S$\&$K(in)& $S_3$& $S_4$&$S_5$&$S_6$\cr
  \hline 

	0.0$\&$0.4&8.95/2.46&74.04/53.18&66.48/36.26&71.98/23.32\cr
	0.1$\&$0.0&44.62/20.51&33.4/12.93&23.63/5.11&16.7/0.65\cr
	0.1$\&$0.9&49.58/31.16&100.00/99.96&99.99/99.91&100.00/99.99\cr
	0.3$\&$0.5&99.92/99.43&99.95/99.75&99.99/99.65&99.96/98.35\cr

  \hline
  \end{tabular}
  \end{center}
\end{table}


\begin{table}
  \begin{center}
  \caption{Power at $95\%$ and $99\%$ confidence level for the $W_k$
statistics (notation used for the table $95\%/99\%$). Results
based on 10000 simulations Gaussian and non Gaussian simulations.
The non Gaussian ones were obtained from the Edgeworth expansion 
for different values of skewness and kurtosis S$\&$K(input).}
  \label{tab2}
  \begin{tabular}{cccccc}\hline
  S$\&$K(in)& $W_1$&$W_2$&$W_3$&$W_4$\cr
  \hline

	0.0$\&$0.4&6.80/1.67&83.03/64.75&77.20/58.76&74.38/52.97\cr
	0.1$\&$0.0&41.13/20.26&33.22/14.31&29.16/12.38&25.61/9.11\cr
	0.1$\&$0.9&53.33/33.90&100.00/100.00&100.00/100.00&100.00/100.00\cr
	0.3$\&$0.5&99.99/99.87&100.00/99.98&99.99/99.97&99.98/99.90\cr

  \hline
  \end{tabular}
  \end{center}
\end{table}


\begin{table}
  \begin{center}
  \caption{Power at $95\%$ and $99\%$ confidence level for the
Shapiro-Francia statistic (notation used for the table 
$95\%/99\%$). Results based on 10000 simulations Gaussian and non 
Gaussian simulations.
The non Gaussian ones were obtained from the Edgeworth expansion 
for different values of skewness and kurtosis S$\&$K(input).}
  \label{tab3}
  \begin{tabular}{cc}\hline
  S$\&$K(in)&SF\cr
  \hline

	0.0$\&$0.4&70.82/47.30\cr
	0.1$\&$0.0&30.97/13.69\cr
	0.1$\&$0.9&100.00/100.00\cr
	0.3$\&$0.5&100.00/99.89\cr
  \hline
  \end{tabular}
  \end{center}
\end{table}


In this paper we have constructed non-Gaussian simulations where the Edgeworth expansion has cumulants of order 5 or higher equal to zero. In \citet{cayon}, it is considered the study of nonGaussian simulations with non zero cumulants of order 5 and 6. The presence of these cumulants seems to be better detected by the Shapiro-Francia test.

\section{MAXIMA Data Analysis. Results}

This analysis has been carried out by \citet{cayon}. Below we briefly summarise the main steps of this analysis.

Suppose we have CMB data in real space. We do not need to have the data on a regular grid. Suppose that these data $x_i$ have mean value $\langle x_i \rangle = 0$, and that the correlation matrix components are given by $C_{ij}=\langle x_i x_j \rangle$, where brackets indicate mean value along several (infinity) realizations. We perform the Cholesky decomposition of the correlation matrix: $C=LL^t$, then $y_i = \sum_j L^{-1}_{ij} x_j$ are uncorrelated and normalized data (zero mean, unit dispersion). Moreover, if the CMB distribution is multinormal then the ${y_i}$ data are independent and Gaussian distributed, with zero mean and unit dipersion. The tests here described are applied to the ${y_i}$ data.

As a real application of this method we analyse the MAXIMA data (Balbi et al., 2000, Hanany et al., 2000). The pixels of the border of the observed region have specially high noise levels. Because of that, we have selected pixels in the central observed area. We have selected pixels with right ascension from 226.47 to 238.24 degs and declination from 226.47 to 238.24 degs. The selected region has 2164 pixels. The selected data is then transformed by multiplying it by the inverse of the Cholesky matrix. Finally, we calculate the above introduced tests. 

After calculation of the statistics we compare these values with the distribution of the Gaussian case. The values of the statistics of the data and the corresponding confidence levels are given in \citet{cayon}. The value of all the statistics used in the goodness-of-fit analysis indicate that the MAXIMA data are compatible with Gaussianity.

We have compared the values of the statistics with distributions of independent $N(0,1)$ data. To be sure that the decorrelation of data does not introduce any artifact that could make this comparison not appropiate, we make the comparison with simulations of MAXIMA, as it is explained in \citet{cayon}. No significant differences are found.

\subsection{Constraints on skewness and kurtosis}

We have found that MAXIMA data are compatible with Gaussianity. Now, we want to constrain the skewness and the kurtosis of MAXIMA data. Firstly, we constrain the values of the statistics $S_3$ and $S_4$. These values are directly related to the skewness and kurtosis (see Section 2.2.1). Since MAXIMA data are Gaussian and the number of data is relatively large, then $S_3 \sim \chi_1^2$ and $S_4 \sim \chi_2^2$ (note that we renormalize the data to zero mean and unit variance, and therefore $S_k \sim \chi_{k-2}^2$ for $k \ge 2$). We denote by $\tilde{S_3}$ and $\tilde{S_4}$ the values of $S_3$ and $S_4$ such that the probability of obtain $S_3 \leqslant \tilde{S_3}$ and $S_4 \leqslant \tilde{S_4}$ is 99$\%$ (if Gaussianity holds). Then, the absolute value of the skewness is equal or smaller than $(6 \tilde{S_3}/n)^{1/2}$ with a probability of 99$\%$ (because of the relation between the skewness and $S_3$). In a similar way, the absolute value of the kurtosis is equal o smaller than $(24 \tilde{S_4}/n)^{1/2}$ with the same probability (this constraint is calculated with the skewness equal to zero). Note that we are working with the transformed independent data $y_i$ of the original correlated MAXIMA data $x_i$. Therefore, the constraints that we have calculated are on the skewness and kurtosis of the $y_i$ data, but we want to constrain these values on the $x_i$ data. These two variables are related by the Cholesky matrix: $x_i = \sum_j L_{ij} y_j$. We denote by $S_y$ and $K_y$ the skewness and kurtosis of the $y_i$ data, and by $S_x$ and $K_x$ the skewness and kurtosis of the $x_i$ data. Then, we have the following relations:

\begin{eqnarray} 
S_x & \equiv & \frac{1}{n} \sum_j \bigg [ \frac{\langle x_j^3 \rangle}{\langle x_j^2 \rangle^{3/2}}\bigg ] = \frac{S_y}{n}\sum_{j} \frac{1}{\langle x_j^2 \rangle^{3/2} }\sum_{i}(L_{ji})^3 ,\nonumber\\ 
K_x & \equiv & \frac{1}{n} \sum_j \bigg [ \frac{\langle x_j^4 \rangle}{\langle x_j^2 \rangle^2} - 3 \bigg ] =\frac{K_y}{n}\sum_{j} \frac{1}{\langle x_j^2 \rangle^2 }\sum_{i}(L_{ji})^4 ,\nonumber
\end{eqnarray}

where $n$ is the number of data and $\langle x_j^2 \rangle = \sum_{i} (L_{ji})^2$. In this way, we calculate the constraints on $S_x$ and $K_x$ once the constraints on $S_y$ and $K_y$ are set. The limits at the 99$\%$ confidence level for the skewness and kurtosis for the MAXIMA experiment are $|S_x| \leqslant 0.035$  and $|K_x| \leqslant 0.036$. 
 
\section{Discussion and conclusions}

Three new methods to test Gaussianity are presented and tested. Non-Gaussian simulations are constructed with the Edgeworth expansion which includes deviations of Gaussian distribution including non zero cumulants of order higher than 2. The statistic $W_2$ developed by \citet{thomas} seems to be the most powerful when there are only cumulants of order 3 and 4. 

A fundamental hypothesis of the methods developed here is that the data must be independent, but the MAXIMA data are dependent because the cosmic signal as well as the instrumental noise are correlated. Then, we decorrelate them with the Cholesky decomposition. In this way, under the Gaussian hypothesis, the data are independent. This decomposition limits the number of data with which we can work, because the number of operations of the Cholesky decomposition and Cholesky matrix inversion is of order $\sim O(n^3)$ (Press et al., 1994). Therefore, the calculations with large $n$ are computationally very expensive. This is a limit to the method here presented.

We have applied the methods to MAXIMA data. These data have been found to be compatible with Gaussianity under these statistical tests (Cay\'on et al., 2003). Constraints on skewness and kurtosis are set to 0.035 and 0.036, respectively. 

In this work we have tested the univariate Gaussian function. In the context of the Rayner and Best test it is possible to analyse directly the multinormal function. In this new approach, a set of ortonormal functions (Rayner and Best, 1990) on the multinormal function is constructed. This will be done in a future work and it will complete the analysis of the univariate Gaussian distribution.





\end{document}